\documentclass[%
 reprint,
 amsmath,amssymb,
 aps,
]{revtex4-1}

\usepackage{graphicx}
\usepackage{dcolumn}
\usepackage{bm}

\begin{document}

\preprint{APS/123-QED}

\title{ Topological phases due to the coexistence of superconductivity and 
spin-density wave: Application to high T$_c$ superconductors in the 
underdoped regime}

\author{Amit Gupta}
\email{sunnyamit31@gmail.com}

\author{Debanand Sa}%
\email{debanandsa@rediffmail.com}
\affiliation{%
Department of Physics, Banaras Hindu University, Varanasi-221 005
  \\}%
  
\date{\today}

\begin{abstract}
 We consider the coexistence of superconductivity(SC) and 
spin-density wave(SDW). The SC is presumed to be of $d_{x^2-y^2}+id_{xy} 
(d_1 + i d_2) $  type whereas the SDW order parameter is of $BCS$/$d_{x y} $ symmetry.
The Hamiltonian having such a structure is shown to have topological coexistence phases in 
addition to the conventional one. It is shown that the amplitudes of both 
the order parameters determine the nature of topological phases. A phase diagram 
characterizing different topological phases with their Chern numbers are 
obtained. The experimental realization of such topological phases with 
reference to high temperature superconductors in the extreme underdoped regime 
are discussed.
\begin{description}
\item[PACS numbers]{74.20.Fg, 74.90.+n, 71.10.Fd, 73.43.-f}
\end{description}
\end{abstract}

\pacs{Valid PACS appear here}
\maketitle



Various quantum states of matter are described by the principle of spontaneous 
symmetry breaking \cite{anderson}. For example, a crystalline solid breaks  
translational symmetry whereas magnetism breaks rotational symmetry and a 
superconductor breaks the more subtle guage symmetry. This class of symmetry 
breaking phenomena are described by an unique order parameter which yields a 
non-vanishing expectation value only in the ordered state. This phenomena is 
described by an effective field theory, known as Landau-Ginzburg theory. 
However, the discovery of quantum Hall effect (QHE) \cite{klitzing} in early 
$ 1980's $  gave rise to a new states of matter which cannot be understood by 
the  above mentioned Landau's symmetry breaking paradigm.  In the quantum Hall 
(QH) state, the bulk of the two-dimensional ($ 2 $D) sample is insulating and 
the electric current is carried only along the edge of the sample. The flow of 
this unidirectional current avoids dissipation and gives rise to quantized 
Hall conductance. The QH state provided the first example of a quantum state 
which is topologically distinct from all states of matter known before. The 
precise quantization of the Hall conductance is explained by the fact that it 
is topological invariant, independent of the material parameters 
\cite{ laughlin, thouless} and cannot change unless the system passes through 
a quantum phase transition.  The QH states belong to a class which explicitly 
breaks time-reversal (TR) symmetry due to the presence of magnetic field. In 
last few years, a new class of topological phase has been theoretically 
predicted in two dimensional systems in presence of time-reversal symmetry  
and experimentally  observed in HgTe quantum wells (QW's). In achieving 
topological insulator, the spin-orbit coupling (SOC) plays an essential role 
\cite{konig, moore, hasan, Qi} . This quantum Hall phase is distinguished from 
a band insulator by a single $ Z_{2} $ invariant. This phase exhibits gapless 
spin-filtered edge-states, which allow for dissipationless transport of charge 
and spin at zero temperature and are protected from weak disorder and 
interactions due to time-reversal symmetry. In 2007, this effects is also 
observed in three-dimensional $(3 D)  $ systems whose surface states are 
spin-polarized $ 2 D $  metals  and are characterized by four  $ Z_{2} $ 
invariants. The topological concept is applied to both insulators and 
superconductors which have full energy gap. It is also worth mentioning that 
one gapped state can not be deformed into another gapped state in a different 
topological class unless a quantum phase transition occurs when the system 
become gapless.
 
In what follows, we consider the coexistence of SC and 
SDW. The SC is considered to be of $ d_1 + i d_2 $  type 
whereas the SDW order parameter is of BCS/$ d_{x y} $ symmetry. Such a Hamiltonian 
is shown to yield topological coexistence phases in 
addition to the conventional one. It is shown that the the amplitudes of both 
the order parameters determine the topological phases. The Chern number and 
hence the nature of the topological phases are determined. A phase diagram 
characterizing different topological phases are obtained. The experimental 
realization of such topological phases are discussed.

High magnetic field measurements in high temperature \cite{krishana} SC have 
not only shown the persistence of SC but also a plateau region. This is 
thought to be due to  development of a small $d_{xy}$ component of the SC 
order parameter to the principal component $d_{x^2-y^2}$. Some other 
experiments in presence of magnetic field \cite{aeppli,khaykovich,chang,haug} 
indicated strong evidence for the onset of translational symmetry breaking 
density wave order. Further, recent angle resolved photo-emission 
spectroscopy (ARPES) experiments done on the deeply underdoped cuprate samples 
which are at the border between the antiferromagnetic (AF) and SC phase have 
revealed a full particle-hole symmetry (PHS) gap for 
Bi$_2$Sr$_2$CaCu$_2$O$_{8+\delta}$ (Bi2212) \cite{tanaka,vishik}, 
La$_{2-x}$Sr$_x$CuO$_4$ (LSCO)\cite{ino,razzoli}, 
Bi$_2$Sr$_{2-x}$La$_x$CuO$_{6+\delta}$ (Bi2201)\cite{peng} and 
Ca$_{2-x}$Na$_x$CuO$_2$Cl$_2$ (NaCCOC)\cite{shen}. Such a gap along the nodal 
direction has been observed in systems whose magnetic and transport properties 
range from AF insulator to SC. This data is in excellent agreement with the 
mixed $d_1 +id_2$ SC along the entire Fermi surface.Thus, it is worth 
addressing the physics of the coexistence phase of SC and SDW with the above 
mentioned symmetries of the order parameters. The mean-field treatment of such 
a coexistence of $ d_1+id_2 $ SC and $ d_{x y} $ SDW in a $ 2 $D  square 
lattice is described by the Hamiltonian  ${\cal H}={\cal H}_{\rm
kin}+{\cal H}_{\rm sc}+{\cal H}_{\rm SDW}$, 
\begin{eqnarray}
{\cal H}_{\rm kin}
&=& 
\sum_{k, \sigma} \epsilon_{k}c^{\dagger}_{k,\sigma}c_{k,\sigma}
\nonumber\\ 
{\cal H}_{\rm sc} &=& \sum_{k}(\bigtriangleup_{k}c^{\dagger}_{k,\uparrow}
c^{\dagger}_{-k,\downarrow} 
\nonumber\\
&& + \bigtriangleup_{k+Q}c^{\dagger}_{k+Q,\uparrow}c^{\dagger}_{-k-Q,\downarrow} 
+ H.C.) 
\nonumber\\
{\cal H}_{\rm SDW} &=& \sum_{k, \sigma}(\sigma M_{k} c^{\dagger}_{k,\sigma}
c_{k+Q,\sigma} + H.C.),    
\end{eqnarray}

\noindent where $c^{\dagger}_{{\bm k}\sigma}$ ($c_{{\bm k}\sigma}$) denotes  
creation (annihilation) operator of the  electron with spin 
$ \sigma=(\uparrow,\downarrow) $  at ${\bm k}=(k_x,k_y)$ and  
$\epsilon_{k}=-2t(\cos{k_{x}}a+\cos{k_{y}}a)-\mu$, 
$\bigtriangleup_{k} = \bigtriangleup_{1,k}+i \bigtriangleup_{2,k} = \Delta_1
(\frac{•\cos{{k_x}a}-\cos{k_{y}}a}{2}) 
+ i \Delta_2(\sin{k_{x}a}\sin{k_{y•}a})$ is 
$d_1+i d_2$-wave SC order parameter, $ M_{k}=\lambda_0\sin{k_{x}a}\sin{k_{y}a}$ 
is the SDW order parameter and  $\mu$ is the chemical potential.  
We express momenta in units of  $ \frac{\pi}{a} $, with '$a$' the lattice 
parameter of the underlying square lattice. The self-consistent equations  
for $\bigtriangleup_{k}$ and $M_{k}$ can be written as, 
$\bigtriangleup_{k}= \sum_{k'}g_{kk'}<c_{-k',\downarrow}c _{k',\uparrow}>$ and 
$M_{k}=\sum_{k', \sigma}\frac{J_{k k'}}{2}\sigma<c^{\dagger}_{k',\sigma}
c_{k'+Q,\sigma}>$, where $g_{k,k'}$ and $J_{k,k'}$  
respectively are the microscopic 
interactions inducing SC and SDW.
In the momentum space, the Hamiltonian can be recasted into
${\cal H}=\sum_{k}
\psi^{\dagger} _{k} 
{\cal H}({k})
\psi_{k}$ 
where the four-component spinor $\psi_{k}$ is,  
$\psi^{\dagger}_{k}=(c_{k\uparrow}^{\dagger},c_{-k\downarrow},
c_{k+Q\uparrow}^{\dagger},c_{-k-Q\downarrow})$   
with $  {\cal H}({k}) $ as, 
\begin{equation}
{\cal H}({k})=
\begin{pmatrix}
\epsilon_{k}-\mu & \bigtriangleup_{k} & M_{k} &  0\\
\bigtriangleup^{*}_{k}& -\epsilon_{k}+\mu & 0 & M_{k} \\
 M_{k} &  0 & \epsilon_{k+Q}-\mu & \bigtriangleup_{k+Q} &\\
0 & M_{k} & \bigtriangleup^{*}_{k+Q} &-\epsilon_{k+Q}+\mu
\end{pmatrix}.
\label{eq:2}
\end{equation}
\noindent Employing the nesting property i.e. $ \epsilon_{k+Q}=-\epsilon_{k} $, 
where $Q=(\pi,\pi)$ is the nesting wave vector. Also, 
$\bigtriangleup_{k+Q}=\bigtriangleup_{1,k+Q}+i \bigtriangleup_{2,k+Q} 
=-\Delta_{1}(\frac{\cos k_x-\cos k_y}{2})+i\Delta_{2}\sin{k_{x}}\sin{k_{y}} 
=-\bigtriangleup_{1,k}+i \bigtriangleup_{2,k} = -\bigtriangleup^{*}_{k}$  and 
$ M_{k+Q} = \lambda_0\sin{k_{x}}\sin{k_{y}} = M_{k} $. Thus, the above Hamiltonian 
matrix reduces to, 
\begin{equation}
{\cal H}({k})=
\begin{pmatrix}
\epsilon_{k}-\mu & \bigtriangleup_{k} & M_{k} &  0\\
\bigtriangleup^{*}_{k}& -\epsilon_{k}+\mu & 0 & M_{k} \\
 M_{k} &  0 & -\epsilon_{k}-\mu & -\bigtriangleup^{*}_{k} &\\
0 & M_{k} & -\bigtriangleup_{k} &\epsilon_{k}+\mu
\end{pmatrix}.
\label{eq:3}
\end{equation}

\begin{figure}
\includegraphics[scale=0.6]{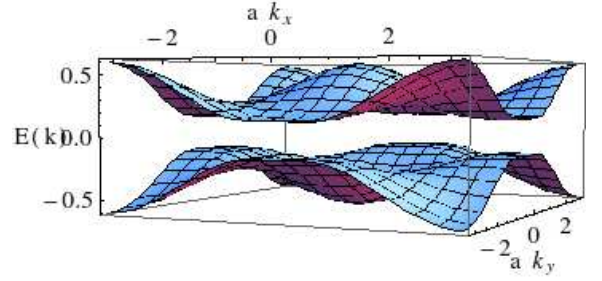} 
\caption{Energy spectra  $ E_{\sigma,+}(k)$, corresponding to 
coexistence of SC order parameter $ d_{x^{2}-y^{2}}+ id_{x y}$ and that of 
the SDW order  parameter (BCS-type) showing fully gapped structure. For illustration, here, we have chosen $ M_0 =.11$ eV, $ \bigtriangleup_{1}=\bigtriangleup_{2}= t$ ($t=0.15$ eV. The chemical potential $ \mu $ is taken to be zero.}
\end{figure}

We will study the phase diagram of the above Hamiltonian for $ \mu =0 $. 
The bulk quasiparticle spectrum is, $ E_{\sigma,\pm}(k) 
=\sigma\sqrt{\epsilon_{k}^{2} +\bigtriangleup_{1,k}^{2} 
+ (\bigtriangleup_{2,k}\pm M_{k})^{2}}$ 
which shows fully gapped structure as shown in 
Fig.1. Considering the case where $<M_k>=0$ but retaining the density-wave 
fluctuations, the above Hamiltonian can be written as, 
\begin{equation} 
{\cal H}({k})=
\begin{pmatrix}
h(k) & 0 \\
0 & -h^{*}(-k)
\end{pmatrix},
\end{equation} 
\noindent where $h(k)=\epsilon_{k}\sigma_{z}
+\bigtriangleup_{1,k}\sigma_x-\bigtriangleup_{2,k}\sigma_y$ 
and $h^{*}(-k)=\epsilon_{k}\sigma_{z}
+\bigtriangleup_{1,k}\sigma_x+\bigtriangleup_{2,k}\sigma_y $  
and $\sigma_x$, $\sigma_y$ and $\sigma_z$ are respectively the Pauli matrices.
The above equation is the Hamiltonian of a standard 
$d_1 +id_2$ superconductor which behaves as a quantum spin Hall 
fluid (QSHF) \cite{volovik,laughlin,senthil} giving rise to spin Hall conductance 
quantization. This Hamiltonian has particle-hole symmetry (PHS)  and a class C with $ SU(2) $ symmetry classification of Altland and Zirnbauer \cite {altland,schnyder} . However, the presence of $ M_{k} $ breaks particle-hole symmetry. 
The topological phases can be characterized by the ${\it Chern}$ 
${\it number}$. 
For a specific model Hamiltonian $h(k)=\sum_{\alpha}d_{\alpha}(k)\sigma_{\alpha}$, 
with $\sigma_{\alpha}$, the Pauli matrices and 
$d_{\alpha}(k)=[d_1(k), d_2(k), d_3(k)]$, 
the Chern number can be calculated from the expression
\begin{equation} 
{\cal N}=\frac{1}{4\pi}\int d^2k\: \hat{d(k)}\cdot(\frac{\partial{\hat d(k)}}
{\partial k_x}\times\frac{\partial {\hat d(k)}}{\partial k_y}), 
\end{equation}

\noindent where the unit vector 
$\hat{d}(k)={\textbf{d}(k)}/\sqrt{\sum d^2(k)}$. Following Volovik \cite{volovik} and Senthil \cite{senthil} the Chern number of the above Hamiltonian is 
calculated as ${\cal {N}}=2\:sgn\:({\Delta_1}{\Delta_2})$ in units of $(\hbar/8\pi)$. 
 This is true for each block of the Hamiltonian $ (4) $. This results remains 
unchanged even if one includes the nesting fluctuations in the   
$d_1+id_2$ superconductor. It is well known from the theory 
of Quantum Hall effect that the quantization of bulk charge/spin Hall 
conductance implies the existence of boundary states carrying charge/spin.  
In the present case the boundary states carry spin. This is due to the 
fact that each block in eqn.(4) can be linearized near the nodal points 
of the square lattice Brillouin Zone. The $d_{x^2-y^2}$ SC part of the 
Hamiltonian makes it Dirac Hamiltonian and $d_{xy}$ part adds a mass 
term to it. From such a structure an effective edge Hamiltonian is 
worked out. Since the edge density operator is proportional to the 
$z$-component of spin, it gives rise to non-zero spin Hall conductance. 

In presence of finite $M_k$, the above Hamiltonian matrix can be written as, 
\begin{equation} 
{\cal H}({k})=
\begin{pmatrix}
h(k) & M_{k} \sigma_0 \\
M_{k}\sigma_0 & -h^{*}(-k)
\end{pmatrix}.
\end{equation}
 
\begin{figure}
\includegraphics[scale=0.6]{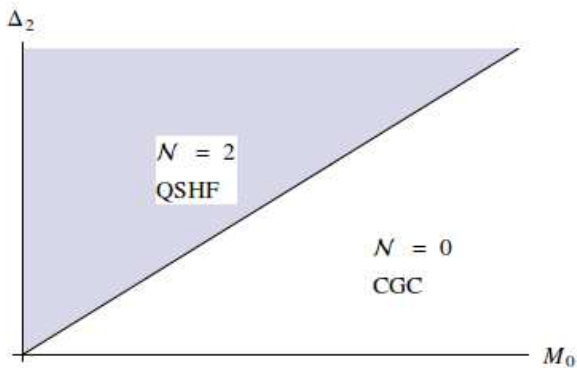} 
\caption{Schematic phase diagram of the coexistence phase of SC $ d_1+i d_2 $ order and 
BCS-type SDW order for $ \mu=0 $ in the case where $k=(\pm\pi/2,\pm\pi/2)$. 
The $x$ axis labels the magnitude of SDW  
order parameter $M_0$  and the $y$ axis labels the magnitude of SC order 
parameter $ \bigtriangleup_2 $. Integer ${\cal N}$ labels the Chern number 
of the coexistence of SC and SDW. }
\end{figure}

The above Hamiltonian (Eqn.(6)) can be block diagonalized through a unitary 
transformation  
\begin{eqnarray}
{\cal H}^{\rm D}({k})=D{\cal H}({k})D^{\dagger},
\quad
D=\frac{1}{\sqrt{2}}
\left(
\begin{array}{cc}
1 & -\sigma_y \\
\sigma_y & 1
\end{array}
\right),  
\end{eqnarray}

\noindent with 
\begin{equation} 
{\cal H}^{\rm D}({k})=
\begin{pmatrix}
{h_{UB}}(k) &0\\
0 & -{h_{LB}}(-k)
\end{pmatrix},
\end{equation} 

\noindent where, ${h_{UB}}(k)=\epsilon_{k}\sigma_{z}
+\bigtriangleup_{1,k}\sigma_x-(\bigtriangleup_{2,k}+M_{k})\sigma_y$ 
and ${h_{LB}}(k)=\epsilon_{k}\sigma_{z}
+\bigtriangleup_{1,k}\sigma_x+(\bigtriangleup_{2,k}-M_{k})\sigma_y $,
It is obvious from the Hamiltonian that the presence of the order parameter 
$M_{k}$ breaks particle-hole symmetry. 

In order to study the phase diagram of this Hamiltonian in the $(\Delta_2,M_0)$ 
plane, one needs to determine the phase boundaries corresponding to gapless 
regions since the topological invariants can not change without closing the 
bulk gap. For the present model, the critical lines are determined by the 
equation $\mid\Delta_2\pm M_0\mid=0$ in the case with $\textbf{k}=(\pm\pi/2,\pm\pi/2)$ which leads to a phase diagram shown in 
Fig. 2. If one considers the SDW order parameter to be of BCS type 
(constant $M_0$), then the line $M_0=\Delta_2$ yields a phase boundary between 
the topological phase (QSHF for $M_0<\Delta_2$) and the conventional gapped 
coexistent (CGC) SDW and SC phase. The Chern number in the case $M_0<\Delta_2$ 
is  ${\cal N}=2$ whereas it vanishes for $M_0>\Delta_2$ till it touches 
$\Delta_2=0$ line. Keeping $M_0$ fixed and varying $\Delta_2$ from zero to 
higher values, one passes from CGC(non-topological) to QSHF(topological) 
phase through a quantum phase transition ($\mid\Delta_2\pm M_0\mid$) line. 
This is due to the fact that on increasing $\Delta_2$, band inversion occurs 
and one enters into the topological phase from an ordinary phase. 
In case of $M_k$ having $d_{xy}$ symmetry, $M_0>\Delta_2$ phase also 
becomes topological which is distinct from $M_0<\Delta_2$ case. The Chern 
number in this case can be inferred from the $\Delta_2=0$ and $M_0\neq 0$ line. 

In order to see the edge state evolution, we have studied the edge states 
numerically on a cylinder geometry with periodic boundary conditions in the 
$y$-direction and open boundary conditions in the $x$-direction as shown in 
Fig 3. The BdG Hamiltonian (Eqn.(3)) has been diagonalized on $N_{x}=80$ sites 
and the energy dispersion $E_k$ verses $k_y$ with edge states has been 
obtained. As already mentioned, two chiral edge states characterize the 
topological QSHF phase.  

\begin{figure}
\includegraphics[scale=0.6]{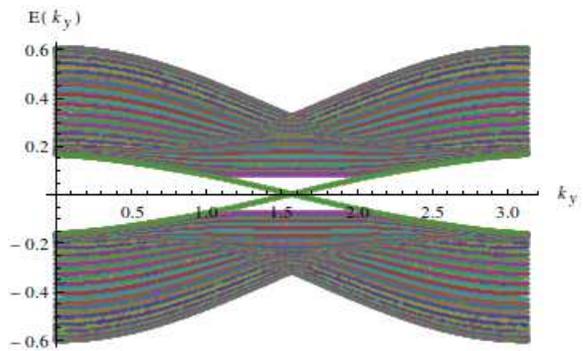} 
\caption{Edge state spectrum of the coexistence phase of SC $ d_1+i d_2 $ order and 
BCS-type SDW order on a cylinder. Parameters are chosen as, $ t=0.15 $ eV,
$M_0 =.075 $ eV  and $ \bigtriangleup_{1}=\bigtriangleup_{2}=t $ for a lattice 
of $ N_{x}=80 $ sites.}
\end{figure}

An intuitive way to understand the topological QSHF state for $M_0=0$ case is 
the evolution of edge states. Such a phase is described by an effective 
one-dimensional Hamiltonian $H_{edge}=\sum_k v\;k\:\eta_k^\dagger\eta_k^\dagger$, 
where $\eta_k$'s are the complex fermion operators. These are
a pair of chiral fermions from each of the four nodal points in the Brillouin 
Zone which give rise to Chern number ${\cal N}=2$. Along the $\Delta\neq 0$ 
and $M_0=0$ line, the $d_1+id_2$ edge states evolve independently. The edge 
states with $k=(\pm\pi/2,\pm\pi/2)$ has width $\xi\sim {\mid\Delta_2\mid}^{-1}$. But in the case of finite $M_0$ ($M_0<\Delta_2$), the widths are 
$\xi_1\sim {\mid\Delta_2-M_0\mid}^{-1} $ and 
$ \xi_2\sim {\mid\Delta_2+M_0\mid}^{-1}$. As $M_0$ increases the localization 
length $\xi$ of the edge modes diverge and gradually they merge into bulk 
states. At the critical line $\mid\Delta_2\pm M_0\mid=0$, these edge states 
completely merge into the bulk states and the system become gapless. 
For $\Delta_2<M_0$, again a gap opens up and the system becomes the CGC phase. 
The present study analyzes the possibility of having topological phases in the 
coexistence phase of SC-SDW, however on the contrary, there exists literature  
either in chiral SC from quantum Hall systems\cite{qi} 
or in topological insulators in a perpendicular magnetic field\cite{zyuzin} 
as well as topological insulating nanowires proximity to SC in longitudinal 
magnetic field\cite{cook}. 

\begin{figure}
\includegraphics[scale=0.7]{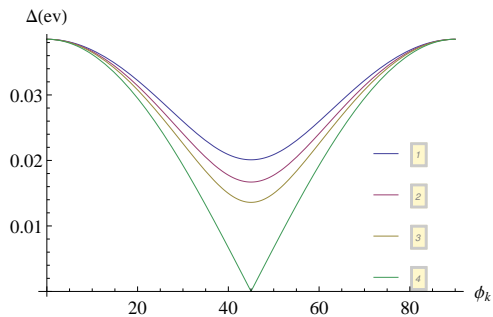}
\caption{Calculated energy gap $\Delta_k=\mid\Delta_{1,k} 
+i(\Delta_{2,k}\pm M_0)\mid$ as a function of Fermi surface angle $\phi_{k}$ where 
$\Delta_1$=38.52 meV and from top to bottom $\Delta_2$ = 20, 16.6, 13.6 
and 0.0 meV respectively consistent with \cite{razzoli}.  }
\end{figure}
  
It has already been known from neutron scattering measurements\cite{lake} 
that there are strong indications of the presence of static/fluctuating 
SDW order below optimal doping at low temperature. The profound departure 
from the $d_{x^2-y^2}$ SC along the nodal direction and the entire gapped Fermi 
surface in the highly underdoped LSCO ($x$=0.08) suggest that the effective 
gap of the present kind, i.e., $\Delta(k)=\mid\Delta_{1,k} 
+i(\Delta_{2,k}\pm M_0)\mid$ might be at work. Considering 
$\Delta_{1,k}=\Delta_1\cos{2\phi_k}$, $\Delta_{2,k}=(\frac{\Delta_2}{2}) 
\sin{2\phi_k}$ one can compute the effective gap $\Delta(\phi_k)$ as a 
function of $\phi_k$. The observed gap \cite{razzoli} 
has a maximum value at the zone boundary($\phi_{k}=0$) and  
decreases monotonically along the Fermi surface to a minimum value at the 
zone diagonal($\phi_{k}=\pi/4$). Below T$_c$(20 K), a finite gap of amplitude 
$\sim$ 20 meV along the zone diagonal has been observed. As temperature 
increases, the diagonal gap monotonically decreases and disappears at 88 K. 
At higher temperature, Fermi arc appears and the arc length increases with 
temperature.  Following \cite{razzoli}, we take $\Delta_1$=38.52 meV to fit 
the data which yields $\Delta_2$=20, 16.6, 13.5, 0 meV 
(from top to bottom in Fig.4) for $M_0/\Delta_2= 10^{-4}$. 
This is in excellent agreement with the experiment data along the entire 
Fermi surface. Recently, there is an attempt \cite{razzoli} to explain the 
data with a pure $d_1+id_2$ SC but the question of whether the SDW ordering 
is present remains open. However, the present study takes into account of SDW ordering along with $d_1+id_2$ SC which explains the data very well for extremely small value of SDW order parameter $ M_{0}$ mentioned above. Further, it unfolds the topological aspects which might be crucial for further understanding of high $ T_ {c}$ SC. Similar approach has been initiated recently for $(p+ip)_{\uparrow\downarrow}$ 
SC in presence of SDW ordering \cite{lu}. \\

In conclusion, we summarize the contents of the paper. Motivated by high magnetic field, neutron scattering and ARPES measurements, we considered a coexistence of $ d+ i d $ SC and BCS like SDW. It has been shown that for SC order parameter ($ \bigtriangleup_{2}$) $ > $ SDW order parameter ($M_{0} $)  the system becomes topological whereas for  $ \bigtriangleup_{2} < M_{0} $ it is conventional coexistence phase. In terms of the amplitude of both the order parameters  $ \bigtriangleup _{2}$ and   $ M_{0} $ a phase diagram characterizing both the phases are obtained. The effective gap is computed for the entire Fermi surface which is in excellent agreement with the observed ARPES data for extremely small value of the SDW order parameter. The present analysis might help to understand the issue of whether the deep underdoped regime in high $ T_{c} $
is characterized by SDW ordering.


\begin{center} 
{\bf ACKNOWLEDGEMENTS}
\end{center} 
  
Financial supports from CSIR, India are gratefully 
acknowledged.  

\bibliography{apssamp}

 
\end{document}